\documentstyle[preprint,aps]{revtex}
\begin{document}
\draft
%
%
%
%
%
%
\title
{
Green Function of the Sutherland Model\\
 with SU(2) internal symmetry} 
\author
{Yusuke Kato\cite{present}}
\address{Department of Physics, Tohoku University, Sendai 980-77, Japan}
\date{Nov. 11, 1996}
\maketitle
\begin{abstract}
We obtain the hole propagator of the Sutherland model with SU(2) internal symmetry for coupling parameter $\beta=1$, which is the simplest nontrivial case. One created hole with spin down breaks into two quasiholes with spin down and one quasihole with spin up. While these elementary excitations are energetically free, the form factor reflects their anyonic character. The expression for arbitrary integer $\beta$ is conjectured. 
\\
\end{abstract}
%
%
There has been a remarkable development in the study of one dimensional models of particles with the inverse-square interactions\cite{Suth}. 
Stimulated by the pioneering work by Simons, Lee, and Al'tshuler\cite{SLA}, the dynamical correlation functions of the Sutherland model for spinless particles were obtained explicitly with the help of various techniques, such as the supermatrix method\cite{HZ}, 2D Yang-Mills theory\cite{Minahan}, and the theory of the Jack symmetric polynomial\cite{Ha,Lesage}. 

The amazing success in the dynamics is, however, limited to the {\it
spinless }case and much less is known for that of the multicomponent
system. 
The multicomponent Sutherland model (MSM)\cite{HaHaldane,Kawakami}
realizes the multicomponent Tomonaga-Luttinger liquid in the simplest
manner. The deeper understanding of the MSM would be helpful to
understand other strongly correlated electron systems, especially {\it
t-J} model\cite{KY} and the singlet factional quantum Hall
effect\cite{Helve}. 
Thus the study on the dynamics of the MSM is highly desirable. 

In this Letter, we present the results on the one-particle Green function of the Sutherland model for particles with SU(2) spin 
($\sigma_i =\pm 1$). This model is the simplest example of the MSM. This is the first calculation of the dynamical correlation function of integrable continuum models with {\it internal symmetry}. 

We consider the following Hamiltonian: \cite{HaHaldane}
\begin{equation}
\hat{\cal H}=-\sum_{i=1}^N \frac{\partial ^2}{\partial
x_i^2}+\frac{2\pi^2}{L^2}\sum_{i<j}\frac{\beta\left(\beta
+P_{ij}\right)}{\sin^2\left[\pi\left(x_i
-x_j\right)/L \right]}\label{hamil}
\end{equation}
for $N$-particle system. Here $P_{ij}$ is the operator that exchanges the spin of particles $i$ and $j$\cite{Polychronakos}. The dimensionless coupling parameter $\beta$ takes non-negative integer values. $L$ is the linear dimension of the system. The statistics of particles are chosen as boson (fermion) for odd (even) $\beta$, so that we can set $P_{ij}=\left(-1\right)^{\beta +1} M_{ij}$ with the coordinate exchange operator $M_{ij}$\cite{Polychronakos}. 

For the model (\ref{hamil}), we calculate the hole propagator: 
\begin{equation}
G^\beta\left(x,t\right)
= \langle
{\rm g},N\vert\hat \psi_\uparrow ^{\dagger}\left(x,t\right)\hat
\psi_\uparrow\left(0,0\right)\vert {\rm g},N\rangle/\langle {\rm g},N\vert {\rm g},N\rangle \label{holepropagator}
\end{equation}
for $\beta=1$, which is the simplest nontrivial case. Here
$\hat\psi_\uparrow(x,t)$ represents ${\rm e}^{{\rm i}\hat{\cal H}t}\hat\psi_\uparrow(x){\rm e}^{-{\rm i}\hat {\cal H}t}$. $|{\rm g},N\rangle$ is the ground state for $N$-particle system. 
The main result of this paper is the following explicit expression in the thermodynamic limit with $\rho_0=N/L$:
\begin{eqnarray}
G^{\beta=1}(x,t)&=& a_1\int_{-1}^{1}{\rm d}v_1\int_{-1}^{1}{\rm d}v_2\int_{-1}^{1}{\rm d}v_\uparrow \frac{\left|v_1 -v_2\right|^{4/3}
\left|v_\uparrow-v_1 \right|^{-2/3}
\left|v_\uparrow-v_2 \right|^{-2/3}
}{\left[\epsilon_1\left(v_1\right)\epsilon_1\left(v_2\right)\epsilon_1\left(v_\uparrow\right)\right]^{1/3}}\nonumber\\
&\times&\exp\left[-{\rm
i}t\left\{\epsilon_1\left(v_1\right)+\epsilon_1\left(v_2\right)+\epsilon_1\left(v_\uparrow\right)-\zeta_1 
\right\}+{\rm
i}\left(v_1+v_2+v_\uparrow \right)\pi x \rho_0/2\right],\label{green1}
\end{eqnarray}
with
$a_1=\rho_0 \zeta_1\Gamma\left[2/3\right]/(72\Gamma^2\left[1/3\right])$, $\zeta_1=\left(3\pi \rho_0/2\right)^2$, and $ \epsilon_1\left(v\right)=3\left(\pi \rho_0/2\right)^2\left(1-v^2\right).
$

In the following paragraphs, we present the derivation of (\ref{green1}). 
In the last part of this paper, we conjecture the expression for arbitrary non-negative integer $\beta$ on the basis of the result for $\beta=1$. 
We consider only the case where $N$ is even and $M=N/2$ is odd.

The ground state is singlet and its wavefunction
$\Phi_{\rm g}^N\left(\left\{X_i\equiv \exp[{\rm i }2\pi x_i/L],\sigma_i\right\}\right)$ is given by \cite{HaHaldane}
\begin{eqnarray}
&\Phi_{\rm g}^N\left(\left\{X_i,\sigma_i\right\}\right)=\prod_{i<j}^N& \left\{\left(X_i -X_j\right) ^{\beta +\delta\left[\sigma_i,\sigma_j\right]}
\left({\rm i}\right)^{(\sigma_i-\sigma_j)/2}\right\}\nonumber\\
&\times\prod_{i=1}^{N}X_i^{-\left\{\beta\left(N-1\right)+M-1\right\}/2}&\sum_S P_S ^N\left[\left\{\sigma_i\right\}\right],\label{groundstate}
\end{eqnarray}
with $
P_S ^N\left[\left\{\sigma_i\right\}\right]=\prod_{i \in S}\delta\left[\sigma_i,1\right]\prod_{j \not\in S}\delta\left[\sigma_j,-1 \right]$. 
The sum with respect to $S$ runs over the subset of $\left\{1,2,3\cdots,N\right\}$ with $M$ elements.
The norm of (\ref{groundstate}) is given by $ _N C_M D(M_1,M_2;\beta)L^N$, with $_N C_M=N!/[M!(N-M)!]$ and 
\begin{equation}
D\left(M_1,M_2;\beta\right)=\prod_{i=1}^{M_1}\int_C\frac{{\rm d}z_i}{{\rm
i}2\pi z_i}\prod_{j=1}^{M_2}\int_C \frac{{\rm d}w_j}{{\rm i}2\pi
w_j}\prod_{i<j}^{M_1}\left|z_i -z_j\right|^{2\beta
+2}\prod_{k<l}^{M_2}\left|w_k -w_l\right|^{2\beta
+2}\prod_{i=1}^{M_1}\prod_{j=1}^{M_2}\left| z_i
-w_j\right|^{2\beta}.\nonumber\\
\end{equation}
Here $C$ is the contour counterclockwise on the unit circle. 
The value of $D\left(M_1,M_2;\beta\right)$ was derived in ref. \cite{norm} as
\begin{equation}
A\left(M_1,M_2;\beta\right)\prod_{k=1}^{M_1}\frac{A\left(k-1,k;\beta\right)}{A\left(k,k-1;\beta\right)}, 
\end{equation}
with 
\begin{equation}
A(J_1,J_2;\beta)=\left(J_2 !\right)\prod_{j=0}^{J_2 -1}\frac{\Gamma\left[\left(\beta+1\right)\left(j+1\right)+J_1\right]}{\Gamma\left[\left(\beta+1\right)j+J_1 +1\right]}\quad\mbox{ for }J_2 \ge J_1 -1.\label{a}
\end{equation}
When $J_2 =1$, the product is taken as unity. 

We consider the action of $\psi$ on the ground state.
The wavefunction of the state $\psi_{\uparrow}(0,0)|{\rm g},N\rangle$ is given by 
\begin{equation}
\sum_{k=1}^N\left(-1\right)^{(\beta +1)k}\Phi_{\rm g}^N\left(\left\{X_i,\sigma_i\right\},X_k=1,\sigma_k=1\right).\label{fix}
\end{equation}
Since particles are identical, we pick out the term $k=N$ in Eq. (\ref{fix}): 
\begin{equation}
\Phi_{\rm g}^N\left(\left\{X_i,\sigma_i\right\};X_N=1,\sigma_N=1\right)=\left(-{\rm i}\right)^M \prod_{i=1}^{N-1}X_i^{-\beta/2}\left(X_i -1 \right)^{\beta+\delta\left[\sigma_i,1\right]}\Phi_{\rm g}^{N-1}\label{action}, 
\end{equation}
where $\Phi_{\rm g}^{N-1}$ represents the ground state
wavefunction with momentum $-\pi(M-1)/L$. 
We calculate the propagator for Eq. (\ref{action}) and then multiply it by $N$. The factor $N$ comes from the number of ways to choose which particle is removed and added. 

For taking the trace over spin variables in Eq. (\ref{holepropagator}), 
we replace the Hamiltonian (\ref{hamil}) by 
\begin{equation}
\hat{\cal H}_\beta=-\sum_{i=1}^N \frac{\partial ^2}{\partial
x_i^2}+\frac{2\pi^2}{L^2}\sum_{i<j}\frac{\beta\left(\beta
-\left(-1\right)^\beta M_{ij}\right)}{\sin^2\left[\pi\left(x_i
-x_j\right)/L \right]}. \label{hamilbeta}
\end{equation}
Spin variable for each particle is then a good quantum number since Hamiltonian (\ref{hamilbeta}) does not contain spin variables. Hence we can fix spin for each particle and multiply the result by $_{N-1} C_{M-1}$ as an equivalent 
procedure to trace over spin variables. It is from this advantage that we consider (\ref{hamilbeta}) instead of (\ref{hamil}). In the following, we set $\sigma_i=1$ for $1\le i \le M-1$ and $\sigma_i =-1$ for $M \le i \le
2M-1$. 

After tracing over $\left\{\sigma_i\right\}$, we arrive at the
following expression:
\begin{eqnarray}
G^{\beta}(x,t)&=&c_0 \prod_{i=1}^{N-1}\left(\oint\frac{{\rm
d}X_i}{{\rm i}2\pi X_i}\right)\overline{\tilde \Phi^{
N-1}_{\rm g}}\prod_{i=1}^{M-1}\left(\bar 
z_i -1\right)^{\beta +1}\prod_{j=1}^{M}\left(\bar w_j
-1\right)^{\beta}\nonumber\\
&\times&{\rm e}^{-{\rm i}\left\{\left(\hat {\cal H}_\beta-E_{\rm g}^{N}\right)t-\hat 
P x\right\}}\prod_{i=1}^{M-1}\left(z_i
-1\right)^{\beta+1}\prod_{j=1}^M\left(w_j -1\right)^{\beta}\tilde \Phi^{N-1}_{\rm g}, 
\label{tildebeta}
\end{eqnarray}
with $c_0 = _{N-1}C~_{M-1}\rho_0/[ _N C_M 
D\left(M_1,M_2,\beta\right)]$. The overlines represent complex conjugate. 
Here $\tilde \Phi_{\rm g}^{N-1}$ is given by
\begin{eqnarray}
\tilde \Phi^{N-1}_{\rm g}&=&\prod_{i<j} ^{M-1}\left(z_i-z_j\right)^{\beta
+1}\prod_{k<l}^M \left(w_k
-w_l\right)^{\beta+1}\prod_{i=1}^{M-1}\prod_{j=1}^{M}\left(z_i
-w_j\right)^\beta \nonumber\\
&\times&\prod_{i=1}^{M-1}z_{i}^{-\left[\beta\left(N-1\right)+M-1\right]/2}
\prod_{j=1}^{M}w_{j}^{-\left[\beta\left(N-1\right)+M-1\right]/2},
\end{eqnarray}
where $z_i$ $\left(1\le i \le M-1\right)=X_i$ and $w_j$ $\left( 1\le j \le
M\right)=X_{j+M-1}$ represent the complex coordinates of particles with spin up
and down, respectively. 

Instead of calculating $G^{\beta=1}(x,t)$ directly, we first calculate the following correlation function for later convenience:
\begin{eqnarray}
\tilde G^{\beta}(x,t)&\equiv&
c_0\prod_{i=1}^{N-1}\left(\oint\frac{{\rm
d}X_i}{{\rm i}2\pi X_i}\right)\overline{\tilde \Phi^{
N-1}_{\rm g}}
\prod_{i=1}^{M-1}\left(\bar 
z_i -1\right)^2\prod_{j=1}^{M}\left(\bar w_j
-1\right)\nonumber\\
&\times&{\rm e}^{-{\rm i}\left\{\left(\hat {\cal H}_\beta-E_{\rm g}^{N}\right)t-\hat 
P x\right\}}
\prod_{i=1}^{M-1}\left(z_i
-1\right)^{2}\prod_{j=1}^M\left(w_j -1\right)\tilde \Phi^{N-1}_{\rm g}.\label{tilde1}
\end{eqnarray}
From (\ref{tildebeta}) to (\ref{tilde1}), we just replace $\prod_{i}\prod_{j}\left(z_i -1\right)^{\beta +1}\left(w_j -1\right)^\beta$ by $\prod_i \prod_j \left(z_i -1\right)^2\left(w_j -1\right)$ but leave the rest of $\beta$ as an arbitrary integer. 

The first step in calculating $\tilde G^\beta(x,t)$ is to expand 
$
\prod_{i=1}^{M-1}\left(z_i
-1\right)^{2}\prod_{j=1}^M\left(w_j -1\right)\tilde \Phi^{N-1}_{\rm g}
$
in terms of eigenfunctions of the Hamiltonian $\hat{\cal H}_\beta$. Relevant eigenstates
have the following form:
\begin{equation}
K\left(\left\{z_i,w_j\right\}\right)\tilde \Phi_{\rm g}^{N-1}\label{eigen}.
\end{equation}
Here $K(\left\{z_i,w_j\right\})$ represents a polynomial satisfying the following
conditions: 
\begin{itemize}
\item symmetric with respect to exchange between  $z_i$'s and
exchange between $w_j$'s 
\item  polynomial of degree not more than 2 with respect to $z_i$ and of degree not more than 1 with respect to $w_j$. 
\end{itemize}

Now we assign two types of indices to eigenstates described by (\ref{eigen}). 
The polynomial part $K\left(\left\{z_i, w_j\right\}\right)$ is written by the linear
combination of $\tilde
m_{\mu}\left(\left\{z_i,w_j\right\}\right)=m_{\mu^{\uparrow}}\left(\left\{z_i\right\}\right)m_{\mu^{\downarrow}}\left(\left\{w_j\right\}\right)$.
Here $m_{\mu^\uparrow}\left(\left\{z_i\right\}\right)$ and
$m_{\mu^\downarrow}\left(\left\{w_j\right\}\right)$ are monomial symmetric functions 
of $\left\{z_i\right\}_{i=1}^{M-1}$ and
$\left\{w_j\right\}_{j=1}^{M}$, respectively. These are given by 
\begin{equation}
m_{\mu^\uparrow}\left(\left\{z_i\right\}\right)=\sum_{P}\prod_{i}^{M-1}z_{P(i)}^{\mu_i^\uparrow},\quad m_{\mu^\downarrow}\left(\left\{w_j\right\}\right)=\sum_{P}\prod_{j}^M w_{P(j)}^{\mu_j^\downarrow},
\end{equation}
respectively. The sum with respect to $P$ runs over distinct permutations. Here $\mu$ is defined by
$\mu=\left\{\mu^\uparrow;\mu^\downarrow\right\}$, where
$\mu^\uparrow=\left\{\mu_1^{\uparrow},\mu_2^\uparrow,\mu_3^\uparrow\cdots 
\mu_{M-1}^\uparrow\right\}$
and
$\mu^\downarrow=\left\{\mu_1^{\downarrow},\mu_2^\downarrow,\mu_3^\downarrow\cdots,\mu_{M}^\downarrow\right\}$ 
are partitions arranged in decreasing order $(\mu_1^\uparrow \ge
\mu_2^\uparrow \ge \mu_3^\uparrow \cdots,\quad \mu_1^\downarrow \ge
\mu_2^\downarrow \ge \mu_3^\downarrow \cdots)$. The weight
$\sum_{i=1}^{M-1}\mu_i ^\uparrow$ $(\sum_{j=1}^M
\mu_j ^\downarrow)$ is denoted by $|\mu^\uparrow|$ ($|\mu^\downarrow|$).

Next we define $\tilde \mu^\uparrow$ and $\tilde \mu^\downarrow$ by
$
\tilde \mu^{\uparrow }=\mu^\uparrow +\left\{M-2,M-3,\cdots,0\right\}
$
and
$
\tilde \mu^{\downarrow }=\mu^\downarrow
+\left\{M-1,M-2,\cdots,0\right\}$, respectively. 
Furthermore we introduce $\mu^+ $ as the rearranged series of  $\tilde
\mu^\uparrow \oplus \tilde \mu^\downarrow $ in decreasing order.
Here we define the order of $\mu^+$ so that $\mu^+ >\nu^+$ if the
first nonvanishing difference $\mu_i^+ -\nu_i^+$ is positive. Next we
define the order of $\mu=\left\{ \mu^\uparrow;\mu^\downarrow\right\}$. If
$\mu^+ >\nu^+$, then
$\mu >\nu$.
If $\mu^+ =\nu^+$ and $|\mu^\downarrow|>|\nu^\downarrow|$, then $\mu > 
\nu$.

We define the polynomial $K_{\mu}\left(\left\{z_i,w_j\right\}\right)$ as 
\begin{itemize}
\item $K_\mu\left(\left\{z_i,w_j\right\}\right)=\tilde
m_\mu\left(\left\{z_i,w_j\right\}\right)+\sum_{\nu\left(<\mu\right)}v_{\mu \nu}\tilde
m_\nu\left(\left\{z_i,w_j\right\}\right)$
\item
$K_\mu\left(\left\{z_i,w_j\right\}\right)\tilde \Phi_{\rm g}^{N-1}
$ is an eigenfunction of $\hat {\cal H}_\beta$.
\item
$K_{\mu}\left(\left\{z_i,w_j\right\}\right)\tilde \Phi_{\rm g}^{N-1}
$'s are
orthogonal with one another.  
\end{itemize}
Eigenfunctions of $\hat {\cal H}_\beta$ are obtained for
distinguishable particles by the same method as used for the single component
bosonic case \cite{Bernard,Katoprl}.  

Although the index $\mu$ seems natural to describe the polynomial
$K_\mu$, here we introduce another index
$\lambda=\left\{\lambda_1,\lambda_2,\lambda_\uparrow\right\}$, which is related directly with rapidities or velocities of elementary excitations. $\lambda_1$ ($\lambda_2$) is the smaller
(larger ) non-negative integer among the set $\left\{M,M-1,\cdots,0\right\}-\tilde
\nu^\uparrow$ and $\lambda_\uparrow$ is among $\left\{M,M-1,\cdots,0\right\}-\tilde
\nu^\downarrow$. In Fig. 1, we illustrate the case of
$\mu=\left\{\mu^\uparrow;\mu^\downarrow\right\}=\left\{2211;1111\right\}$ for example. 
We see that there are three vacant rapidities
$\lambda=\left\{M-5,M-1,M-4\right\}$. Now we identify $\lambda_1,\lambda_2$ and
$\lambda_\uparrow$ as rapidities of {\it quasiholes} with spin down and up, 
respectively. 

In order to calculate $\tilde G^\beta$, we have to know the expansion
coefficient $C_\lambda$ which appears in 
\begin{equation}
\prod_{i=1}^{M-1}\left(z_i -1\right)^2 \prod_{j=1}^{M}\left(w_j
-1\right)=\sum_{0\le \lambda_1 <\lambda_2}\sum_{1\le \lambda_2 \le
M}\sum_{0\le \lambda_\uparrow \le M}C_\lambda K_\lambda
\end{equation}
and the norm $N_\lambda$ for each wavefunction:
\begin{equation}
N_\lambda=\prod_{i=1}^{N-1}\oint \frac{{\rm d}X_i}{{\rm i}2\pi X_i}\left|K_\lambda
\tilde \Phi^{N-1}\right|^2.
\end{equation}
In terms of $C_\lambda$ and $N_\lambda$, the hole propagator is
written as
\begin{equation}
\tilde G^{\beta}(x,t)=c_0\sum_{\lambda_1,\lambda_2,\lambda_\uparrow}C_\lambda ^2 N_\lambda {\rm exp}\left[-{\rm i}\left(E_\lambda^{N-1}-E_{\rm g}^{N}\right)t+{\rm i}P_\lambda x\right].\label{hole2}
\end{equation}
Now we present the explicit forms of $E_{\lambda}^{N-1}$, $P_\lambda$, $C_\lambda$, and $N_\lambda$. 
Momentum of the state $\lambda$ is given by
\begin{equation}
P_\lambda =\pi\left(5M
-1\right)/L-\left(2\pi/L\right)\left(\lambda_1 +\lambda_2
+\lambda_\uparrow\right)+\pi\beta\left(1 -2M\right)/L.
\end{equation}
%
In terms of $\lambda=\left\{\lambda_1,\lambda_2,\lambda_\uparrow\right\}$, we write the expression $E_\lambda^{N-1}-E_{\rm g}^N$ as
\begin{eqnarray}
& &
E_\lambda ^{N-1}-E_{\rm g}^N=\nonumber\\
& &\left(2\pi/L\right)^2\left[-\left(M-1/2\right)M\beta^2-\left(M^2 -10M +1\right)/4-\beta \left(2M^2+7M-1\right)/2+\left(2M-1\right)\beta^2/4\right.\nonumber\\& &-\left.\left(1+2\beta\right)\left(\lambda_1 ^2 +\lambda_2 ^2 +\lambda_\uparrow^2\right)+\left(M+2\beta M+\beta-1\right)\left(\lambda_1 +\lambda_2
+\lambda_\uparrow\right)-2\beta\left(\tilde \lambda_1-\tilde \lambda_3\right)\right].
\end{eqnarray}
Here $\tilde \lambda_1$, $\tilde \lambda_2$, and $\tilde \lambda_3$ are the smallest, the second smallest and the largest of $\left\{\lambda_1,\lambda_2,\lambda_\uparrow\right\}$, respectively. 

General expressions of $C_\lambda$ and $N_\lambda$ for the Jack
polynomial have been derived in mathematical literature\cite{Macdonald}. 
In the case of $K_\lambda$, however, these
expressions have not yet been derived. Hence we calculate $K_{\lambda}$ 
 for system with finite number of particles ($M\le 11$) up to degree $\left|\mu\right|=20 $. Based on these data, we obtain the general form of 
$C_\lambda$ and $N_\lambda$. 
The coefficient $C_\lambda$ is given by
\begin{eqnarray}
&C_{\lambda}=&\Gamma^2 \left[\left(1+\beta\right)\eta\right]\left/(\Gamma\left[\left(2+\beta\right)\eta\right]\Gamma\left[\eta\right]\right) 
\nonumber\\
&\times& \left\{
\begin{array}{l}
\frac{\Gamma\left[\lambda_2 -\lambda_1 +\eta\right]}{\Gamma\left[\lambda_2 -\lambda_1 -\beta\eta\right]}
\frac{\Gamma\left[\lambda_\uparrow-\lambda_1 +\left(1-\beta\right)\eta\right]}{\Gamma\left[\lambda_\uparrow-\lambda_1 +\eta\right]}
\frac{\Gamma\left[\lambda_\uparrow -\lambda_2 +\eta\right]}{\Gamma\left[\lambda_\uparrow -\lambda_2 +\left(1+\beta\right)\eta\right]}\\
\mbox{ for }\lambda_1 <\lambda_2 \le \lambda_\uparrow \\
\frac{\Gamma\left[\lambda_2 -\lambda_1 +\left(1-\beta\right)\eta\right]}{\Gamma\left[\lambda_2 -\lambda_1 -2\beta\eta\right]}
\frac{\Gamma\left[\lambda_\uparrow-\lambda_1 +\eta\right]}{\Gamma\left[\lambda_\uparrow-\lambda_1 +\left(1+\beta\right)\eta\right]}
\frac{\Gamma\left[\lambda_2-\lambda_\uparrow -2\beta\eta\right]}{\Gamma\left[\lambda_2-\lambda_\uparrow-\beta\eta\right]}\\
\mbox{ for }\lambda_1 \le \lambda_\uparrow < \lambda_2 \\
\frac{\Gamma\left[\lambda_2 -\lambda_1 +\eta\right]}{\Gamma\left[\lambda_2 -\lambda_1 -\beta\eta\right]}
\frac{\Gamma\left[\lambda_1-\lambda_\uparrow -2\beta\eta\right]}{\Gamma\left[\lambda_1-\lambda_\uparrow-\beta\eta\right]}
\frac{\Gamma\left[\lambda_2-\lambda_\uparrow -3\beta\eta\right]}{\Gamma\left[\lambda_2-\lambda_\uparrow-2\beta\eta\right]}\\
\mbox{ for }\lambda_\uparrow < \lambda_1 < \lambda_2,  
\end{array}
\right.
\end{eqnarray}
with $\eta=1/(1+2\beta)$. 
The norm is given by
\begin{equation}
N_{\lambda}=\frac{D(M-1,M;\beta)\Gamma\left[M\right]\Gamma\left[M+1-\beta\eta\right]
{\cal 
A}_\lambda {\cal B}_\lambda,}{\Gamma\left[M
-\beta\eta\right]\Gamma\left[M+1-3\beta\eta\right]\Gamma^3\left[\left(1+\beta\right)\eta\right]}, 
\end{equation}
with  
\begin{equation}
{\cal A}_\lambda=\prod_{k=1}^3
\frac{\Gamma\left[\tilde \lambda_k +1-k\beta\eta\right]
\Gamma\left[M-\tilde \lambda_k +(1+(k-2)\beta)\eta\right]
}
     {\Gamma\left[\tilde \lambda_k +1+(1-k)\beta\eta\right]\Gamma\left[M-\tilde \lambda_k +(1+(k-1)\beta)\eta\right]}
\end{equation}
and 
\begin{equation}
{\cal B}_\lambda=\left\{
\begin{array}{l}
\frac{\Gamma\left[\lambda_2 -\lambda_1 -\beta\eta\right]
      \Gamma\left[\lambda_2 -\lambda_1 +\left(1+\beta\right)\eta\right]
\Gamma^2\left[\lambda_\uparrow -\lambda_1 +1-2\beta\eta\right]\Gamma^2\left[\lambda_\uparrow-\lambda_2 +1 -\beta\eta\right]
}
     {
\Gamma\left[\lambda_2 -\lambda_1 \right]
      \Gamma\left[\lambda_2 -\lambda_1 +\eta\right]
\Gamma\left[\lambda_\uparrow-\lambda_1 +1-3\beta\eta\right]
      \Gamma\left[\lambda_\uparrow-\lambda_1 +1 -\beta\eta\right]
     \Gamma  \left[\lambda_\uparrow-\lambda_2 +1-2\beta\eta\right]
      \Gamma\left[\lambda_\uparrow -\lambda_2 +1\right]}
\quad\mbox{ for } \lambda_1 <\lambda_2 \le \lambda_\uparrow \\
   \\
\frac{\Gamma\left[\lambda_2 -\lambda_1 +\eta\right]
      \Gamma\left[\lambda_2 -\lambda_1 -2\beta\eta\right]
\Gamma^2\left[\lambda_\uparrow -\lambda_1 +\left(1+\beta\right)\eta\right]
\Gamma^2\left[\lambda_2-\lambda_\uparrow  -\beta\eta\right]
}
     {\Gamma\left[\lambda_2 -\lambda_1 +\left(1-\beta\right)\eta\right]
      \Gamma\left[\lambda_2 -\lambda_1 -\beta\eta\right]
\Gamma  \left[\lambda_\uparrow -\lambda_1 +1\right]
      \Gamma  \left[\lambda_\uparrow -\lambda_1 +\eta\right]
\Gamma  \left[\lambda_2-\lambda_\uparrow -2\beta\eta\right]
      \Gamma  \left[\lambda_2-\lambda_\uparrow  \right]}\quad\mbox{ for } \lambda_1 \le \lambda_\uparrow < \lambda_2 \\
                                                 \\
\frac{\Gamma\left[\lambda_2 -\lambda_1 -\beta\eta\right]
      \Gamma\left[\lambda_2 -\lambda_1 +\left(1+\beta\right)\eta\right]
     \Gamma^2\left[\lambda_1-\lambda_\uparrow -\beta\eta\right]\Gamma^2\left[\lambda_2-\lambda_\uparrow-2\beta\eta\right]}
     {\Gamma\left[\lambda_2 -\lambda_1 \right]
      \Gamma\left[\lambda_2 -\lambda_1 +\eta\right]\Gamma  \left[\lambda_1-\lambda_\uparrow \right]
      \Gamma  \left[\lambda_1-\lambda_\uparrow -2\beta\eta\right]\Gamma  \left[\lambda_2-\lambda_\uparrow- \beta\eta\right]
      \Gamma  \left[\lambda_2-\lambda_\uparrow-3\beta\eta\right]}\mbox{ for } \lambda_\uparrow <\lambda_1 < \lambda_2. \\
\end{array}\right.
\end{equation}

Now we consider the thermodynamic limit $M \rightarrow \infty$, $L \rightarrow \infty$ with $\rho_0=2M/L$. 
For $\lambda_1$, $\lambda_2$, $\lambda_\uparrow$, $\left|\lambda_1-\lambda_2\right|$, $\left|\lambda_1 -\lambda_\uparrow\right|$, $\left|\lambda_2 -\lambda_\uparrow\right| \sim {\cal O}(M)$, we obtain the asymptotic forms of $C_\lambda$, $N_\lambda$, $E_\lambda^{N-1} -E_{\rm g}^N$, and $P_\lambda$. In the case of $C_\lambda$ and $N_\lambda$, we can use the Stirling formula: $\Gamma\left[N+1\right]\rightarrow \sqrt{2\pi}N^{N+1/2}{\rm e}^{-N}$. The sums with respect to $\lambda_1,\lambda_2$, and $\lambda_\uparrow$ turn into the multiple integral:
\begin{equation}
\left(\frac{2}{M}\right)^3\sum_{\lambda_1=0}^{\lambda_2 -1}\sum_{\lambda_2 =1}^{M}\sum_{\lambda_\uparrow =0}^{M}\rightarrow 
\frac12\int_{-1}^{1}{\rm d}v_1\int_{-1}^{1}{\rm d}v_2\int_{-1}^{1}{\rm d}v_\uparrow,
\end{equation}
with $v_{i}=1-\left(2\lambda_i/M\right)\quad\left(i=1,2,\uparrow\right)$.
From the above results, $\tilde G^{\beta}(x,t)$ reduces to
\begin{eqnarray}
\tilde G^{\beta}(x,t)&=& a_\beta\int_{-1}^{1}{\rm d}v_1\int_{-1}^{1}{\rm d}v_2\int_{-1}^{1}{\rm d}v_\uparrow \frac{\left|v_1 -v_2\right|^{2\left(1+\beta\right)\eta}
\left|v_\uparrow -v_1\right|^{-2\beta\eta}
\left|v_\uparrow -v_2\right|^{-2\beta\eta}
}{\left[\epsilon_\beta\left(v_1\right)\epsilon_\beta\left(v_2\right)\epsilon_\beta\left(v_\uparrow\right)\right]^{\beta\eta}}\nonumber\\
&\times&\exp\left[-{\rm
i}t\left\{\epsilon_\beta\left(v_1\right)+\epsilon_\beta\left(v_2\right)+\epsilon_\beta\left(v_\uparrow\right)-\zeta_\beta 
\right\}+{\rm
i}\left(2-2\beta+v_1+v_2+v_\uparrow \right)\pi x \rho_0/2\right], \label{greenbeta}
\end{eqnarray}
with the prefactor 
\begin{equation}
a_\beta=\frac{\Gamma\left[\left(1+\beta\right)\eta\right]\rho_0  (\eta\zeta_\beta )^{3\beta\eta}M^\gamma}{2^{1+6\eta}\left(2+\beta\right)^{\beta}\Gamma^2\left[\left(2+\beta\right)\eta\right]\Gamma^2\left[\eta\right]}, 
\end{equation}
the chemical potential $\zeta_\beta=[\pi (1+2\beta)\rho_0/2]^2$, the quasihole energy $ \epsilon_\beta\left(v\right)=\left(\pi \rho_0/2\right)^2\left(1+2\beta\right)\left(1-v^2\right)$, and the exponent $\gamma=(5+\beta)\eta -(1+\beta)$.
By setting $\beta=1$ in Eq. (\ref{greenbeta}), we obtain the expression (\ref{green1}). 

Now we discuss the physical implication of (\ref{greenbeta}). We can see that the three quasiholes are energetically free while the form factor has nontrivial structure. From the study of thermodynamics\cite{fesmsm}, on the other hand, we have learned that quasiholes are free but obey nontrivial (exclusion) statistics. Here we attribute the nontrivial form factor to the exclusion statistics. Namely we read the form factor in (\ref{greenbeta}) as
\begin{equation}
\frac{\left|v_1
-v_2\right|^{2g_{\downarrow \downarrow}}
\left|v_\uparrow -v_1\right|^{2 g_{\uparrow \downarrow}}
\left|v_\uparrow -v_2\right|^{2 g_{\uparrow \downarrow}}
}{\left[\epsilon_\beta\left(v_1\right)\epsilon_\beta\left(v_2\right)\right]^{1-g_{\downarrow 
\downarrow}}\left[\epsilon_\beta\left(v_\uparrow\right)\right]^{1-g_{\uparrow 
\uparrow}}}, \label{interpretation}
\end{equation}
where the matrix 
\begin{equation}
g_{\rm h}=\left(
\begin{array}{cc}
g_{\uparrow \uparrow }&g_{\uparrow \downarrow}\\
g_{\downarrow \uparrow}&g_{\downarrow \downarrow}
\end{array}
\right)=\frac{1}{1+2\beta}\left(
\begin{array}{cc}
\beta+1&-\beta\\
-\beta&\beta+1
\end{array}\right)\label{gh}
\end{equation}
represents the statistical interaction among quasiholes of the Sutherland model with SU(2) internal symmetry. Actually we meet the matrix (\ref{gh}) in the thermodynamics of the present model. 

Lastly we consider the hole propagator $G^\beta (x,t)$ with arbitrary
integer coupling $\beta$. In this case, we know that one created hole
with spin down breaks into $\beta+1$
quasiholes with spin down and $\beta$ quasiholes with spin up. Energies
and momenta of intermediate states are obtained easily. In the
thermodynamic limit, we obtain 
$
E_\lambda ^{N-1}-E_{\rm g}^{N}\rightarrow
\sum_{i=1}^{\beta+1}\epsilon_{\beta}\left(v_i\right)
+\sum_{j=1}^{\beta}\epsilon_\beta \left(u_j\right)-\zeta_\beta
$ 
and
$
P_\lambda ^{N-1}\rightarrow\left(\pi \rho_0/2\right)\left(\sum_{i=1}^{\beta +1}v_i
+\sum_{j=1}^{\beta} u_j \right),
$
where $v_i$ ($1\le i\le \beta +1$) and $u_j$ ($1\le j \le \beta$) are
normalized velocities of quasiholes with spin down and up,
respectively. Although we do not have the expression for the form
factor, we make a conjecture relying on the interpretation (\ref{interpretation}). The resultant expression for
$G^\beta(x,t)$ is given by
\begin{eqnarray}
G^{\beta}(x,t)&\sim& \prod_{i=1}^{\beta+1}\int_{-1}^{1}{\rm
d}v_{ i}\prod_{j=1}^\beta \int_{-1}^{1}{\rm d}u_{ j}
 \frac{\prod_{i<j}^{\beta+1}\left|v_{ i} -v_{j}\right|^{2
g_{\downarrow \downarrow }}
\prod_{k<l}^\beta \left|u_{ k} -u_{l}\right|^{2g_{\uparrow \uparrow}}
\prod_{i=1}^{\beta +1}\prod_{j=1}^{\beta}\left|v_{i}
-u_{ j}\right|^{2g_{\downarrow \uparrow }}
}{\prod_{i=1}^{\beta+1}\left[\epsilon_\beta\left(v_{i}\right)\right]^{1-g_{\downarrow 
\downarrow
}}\prod_{j=1}^{\beta}\left[\epsilon_\beta\left(u_{j}\right)\right]^{1-g_{\uparrow 
\uparrow }}}\nonumber\\
&\times&\exp\left[-{\rm
i}t\left\{\sum_{i=1}^{\beta+1}\epsilon_\beta\left(v_{i}\right)+\sum_{j=1}^{\beta}\epsilon_\beta\left(u_{j}\right)-\zeta_\beta\right\}+{\rm
i}\left(\sum_{i=1}^{\beta+1}v_{i}+\sum_{j=1}^{\beta}u_{j}\right)\pi x
\rho_0/2\right], \label{general}
\end{eqnarray}
up to a constant factor. 

The expression (\ref{general}) would be plausible in view of the known form factor in the single component model (with integer coupling $\beta$), which is given by\cite{Minahan,Ha,Lesage}
\begin{equation}
\frac{\prod_{i<j}^{\beta}\left|v_i -v_j\right|^{2g}}{\prod_i^\beta \left[\epsilon(v_i)\right]^{1-g}}.\label{single}
\end{equation}
Here $v_i$ is the velocity and $\epsilon(v)$ is the energy of quasihole. The statistical parameter $g$ was derived as $1/\beta$ from the study of the thermodynamics\cite{BernardWu}. We can see that the form factor in (\ref{general}) is a natural generalization of (\ref{single}). The conjecture (\ref{general}) should be checked microscopically, {\it e.g.} with the use of the theory of the nonsymmetric Jack polynomial\cite{opdam}. This is left as a future work. 

The author thanks Y. Kuramoto and T. Yamamoto for useful comments. 
This work was supported by a Grant-in-Aid for Encouragement of Young Scientists (08740306) from the Ministry of Education, Science, Sports and Culture of Japan.%

\begin{figure}
\caption{A diagram that contributes to $\tilde G^{\beta}$. This figure
corresponds to the case of $\mu=\left\{2211;1111\right\}$. The open
and shaded blocks represent the ground state and an excitated state,
respectively. There are two quasiholes with spin down at $\lambda_1$
and $\lambda_2$, while one with spin up is located at $\lambda_{\uparrow}$.}
\end{figure}

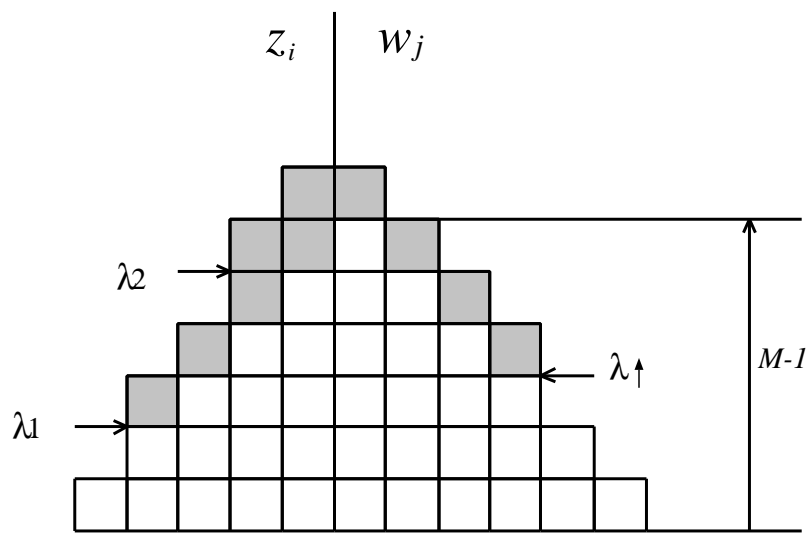
\end{document}